\documentstyle[epsf,prd,aps,floats]{revtex}

\input{psfig.tex}
\tighten
\begin{document} 
\draft
\twocolumn[\hsize\textwidth\columnwidth\hsize\csname@twocolumnfalse\endcsname
\title{A time varying speed of light as a solution to
cosmological puzzles}
\author{Andreas Albrecht and Jo\~ao Magueijo}
\address{Theoretical Physics, The Blackett Laboratory,
Imperial College, Prince Consort Road, London, SW7 2BZ, U.K.}

\maketitle
\begin{abstract}
We consider the cosmological implications of light travelling faster 
in the early Universe. We propose a prescription for
deriving corrections to the cosmological evolution equations while the
speed of light $c$ is changing. We then show how the horizon, flatness,
and cosmological constant problems may be solved. We also study cosmological
perturbations in this scenario and show how one may solve the homogeneity
and isotropy problems. As it stands, our scenario appears to most
easily produce extreme homogeneity, requiring structure to be produced
in the Standard Big Bang epoch.  Producing significant perturbations
during the earlier epoch would require a rather careful design of the
function $c(t)$.  The large entropy inside the horizon nowadays can
also be accounted for in this scenario.
\end{abstract}

\date{\today}

]

\renewcommand{\thefootnote}{\arabic{footnote}}
\setcounter{footnote}{0}

\section{The puzzles of the Big Bang model}

Cosmologists have long been dissatisfied with the ``Standard Big
Bang'' (SBB) model of the Universe.  This is not due to any conflict
between the big bang theory and observations, but because of the limited
scope offered by the SBB to explain certain striking features
of the Universe.  From the SBB perspective the homogeneity, isotropy,
and ``flatness'' of the Universe, 
and the primordial seeds of galaxies and other structure
are all features which are ``built in'' from the beginning as 
initial conditions.  Cosmologists would like to explain these features
as being the result of calculable physical processes.  A great
attraction of the Inflationary Cosmologies \cite{infl} 
is that they address these
issues by showing on the basis of concrete calculations that a wide
variety of initial conditions evolve, during a period of cosmic
inflation, to reflect the homogeneity, isotropy, flatness and
perturbation spectrum that we observe today.  

So far, {\em all} attempts to achieve this kind of 
improvement over the SBB have
wound up taking the basic inflationary form, where the observable
Universe experiences a period of
``superluminal''  expansion.  This is accomplished by modifying the
matter content of the Universe  in such a way  that 
ordinary Einstein gravity becomes repulsive and drives inflationary
expansion. 
This process is 
in many ways remarkably straightforward and has found numerous
realizations over the years (\cite{infl1,infl2b,infl2,infl3}, etc), 
although it might still be argued that a
truly compelling microscopic foundation for inflation has yet to
emerge.   

One interesting question is whether inflation is the {\em right}
solution to the cosmological puzzles.  Is inflation really what nature
has chosen to do?  When this matter is discussed there is a
notable absence of any real competition to inflation, and this must be
counted in inflation's favour.  
However, we believe the picture would become
much clearer if some kind of debate along these lines were possible.
To this end, we discuss here a possible alternative to inflationary
cosmology which, while not as well 
developed as today's inflationary models,
might lead to some illuminating discussion.  

In this alternative
picture, rather than changing the matter content of the Universe, we
change the speed of light in the early Universe.  
We assume that the Universe matter content is the same as in the 
SBB, that is, the Universe is radiation dominated at early times.
We also assume that Einstein's gravity is left unchanged,
in a sense made precise in Section~\ref{post}. The geometry and
expansion factor of the Universe are therefore the same as in
the SBB. However the {\it local} speed of light, as measured
by free falling observers associated with the cosmic expansion,
varies in time, decelerating from a very large value to its
current value.

We discuss below how Varying Speed of Light (VSL) models might 
resolve the same cosmological puzzles as inflation, and 
offer a resolution to the cosmological constant problem as well.
We shall not dwell on the possible mechanisms by means of which
the speed of light could have changed. Rather we wish to concentrate
on the conditions one should impose on VSL models for their cosmological
implications to be interesting. This phenomenological approach 
should be regarded as a curiosity, which, we hope, will prompt
further work towards an actual theory in which the physical basis of
VSL models is realized. 

One may doubt that such a self-consistent theory could ever be
constructed. We therefore feel forced to transcend the scope
of this paper, and discuss essential aspects of such a theory. 
We find it befitting to start our discussion with an assessment
of the experimental meaning of a varying $c$ (Section~\ref{mean}).
We also need to be more specific about VSL theories in order
to tackle the flatness, cosmological constant, homogeneity, 
and entropy problems. In Section~\ref{post} we state
what is actually required from any VSL theory to solve these problems.
However in Appendix I we lay out the foundations for such a theory. 

\section{The meaning of a variable speed of light}\label{mean}
We first address the question of the meaning of a varying
speed of light. Could such a phenomenon be proved or disproved
by experiment? {\it Physically}
it does not make sense to talk about constancy 
{\it or} variability of any dimensional ``constant''. A measurement of
a dimensional quantity must always represent its ratio to some
standard unit. For example, the length of my arm in meters is really
the dimensionless quantity given by the ratio of the arm length to
the length of a meter stick.  If the ratio varied, one {\em could} interpret
this as a variation in either (or both) of the two lengths.
In familiar situations, there is usually a preferred interpretation
which distinguishes itself by giving a simpler view of the
world. Choosing a given person's arm as a standard of length would
require a whole range of simple objects to undergo peculiar dynamics,
whereas assuming the meter stick to be constant would usually give a
much simpler picture.

None the less, a given theory of the world requires dimensional
parameters.  If these parameters varied, how would that process show up in
experiments?  Suppose we set out to measure the speed of light.
For this one needs a length measure (rod) and a clock.  In a world
described by a theory with time varying dimensional parameters, it is
quite possible that the rods and clocks, as well as the photon speeds,
could all vary.  Because measurements are fundamentally dimensionless,
the experimental result will only measure some dimensionless
combination of the fundamental constants.  Let us sketch a simple
illustration:  Suppose we measure time with an atomic clock.  Taking
the Rydberg energy ($E_R = m_e e^4/ 2(4\pi \epsilon_o
)^2\hbar^2$) to represent the dependence of all atomic energy levels on the
fundamental constants, the oscillation period of the atomic clock will be
$\propto  \hbar /E_R$.  Likewise, taking the Bohr radius ($a_0 =
4\pi\epsilon_0\hbar^2 / m_e e^2$) to reflect the 
relationship between the lengths of ordinary objects (made of atoms)
and the fundamental constants, the length of our rod is $\propto a_0$.
Thus a measurement of $c$ with our equipment is really a measurement
of the dimensionless quantity
\begin{equation}
{c \over {a_0  / (\hbar /E_R)}} = {8\pi\epsilon_0\over \alpha}  
\end{equation}
essentially the fine structure constant.
We could of course use other equipment which depends in
different ways on the fundamental dimensionless constants.  For
example, pendulum clocks will necessarily involve Newton's constant
$G$.  Different experiments will result, which measure different
dimensionless combinations of the fundamental dimensional constants. 
Our conclusion that physical experiments are only sensitive to
dimensionless combinations of dimensional constants is hardly a new one.
This idea has been often stressed by Dicke (eg. \cite{dicke}), and we
believe this is not controversial. 

Thus, speaking in theoretical terms of time varying dimensional
constants can lead to problems.
To give an historical example, 
papers \cite{baum,sol} were written claiming stringent 
experimental upper bounds
on the time variability of the dimensional quantity $\hbar c$. 
In these the product $E\lambda$ was found to be the same for light 
emitted at very different redshifts. From the deBroglie relation
$\hbar c=E\lambda$ one infers the constancy of $\hbar c$.
Bekenstein gives an illuminating discussion
of the fallacy built into this argument
\cite{beck}. Built into $E\propto 1/a$ 
and $\lambda\propto a$ is the assumption that $\hbar c$ is constant,
for otherwise the wavevector $k^\mu$ and the momentum vector $p^\mu$
could not both be parallel transported. Hence the experimental
statement that $\hbar c$ is constant is circular. 

What would we do therefore if we were to observe changing dimensionless
quantities? Any theory explaining the phenomenon would necessarily have 
to make use of dimensional quantities.  It would a priori be a matter of 
choice, prejudice, or convenience to decide which dimensional quantities 
are variable and which are constant (as we mentioned in the
illustration above).  There would be a kind of equivalence, or duality between
theories based on any two choices as far as dimensionless observations
are concerned. However, the equations for two theories which are 
observationally equivalent, but which have different dimensional
parameters varying, will in general not look the same, and again
simplicity will end up being an important factor in making a choice
between theories.  In what follows, we will prefer to work with models
which have the simplicity of ``minimal coupling''.

Let us illustrate this point with a topical example. 
There has been a recent claim \cite{webb} of experimental evidence 
for a time changing fine structure constant $\alpha=e^2/(4\pi \hbar c)$.
Although the ongoing chase for systematics precludes 
any definitive conclusions,
let us assume for the purpose of the argument that the effect is real.

In building a theory which explains a variable $\alpha$
we must make a decision. We could {\it postulate} that electric charge
changes in time, or, say, that $\hbar c$ must change in time. 
Bekenstein \cite{bek2} constructs a theory based on the first alternative. 
He postulates a Lorentz invariant action, which does not conserve 
electric charge. 
Our theory is based on the second choice. We postulate breaking
Lorentz invariance, a changing $\hbar c$,  and consequently 
non-conservation of energy. Any arguments against
the experimental meaning of a changing $c$ can also be directed 
at Bekensteins' changing $e$ 
theory, and such arguments are in both cases meaningless. In both cases 
the choice of a changing dimensional ``constant'' reverts to the postulates
of the theory and is not, a priori, an experimental issue.  The
observables are always dimensionless.
However, the {\em minimally coupled} theories based on either choice are
{\em not} dual (as we shall point out in Appendix I). 
For this reason one might prefer
one formulation over the other.

Finally, and on a different tone,
suppose that future experiments were to confirm that not only $\alpha$
changes in time, but also that there are
time variations in dimensionless coupling constants based on other
interactions, 
$\alpha_i=g_i^2/(\hbar c)$\footnote{In writing 
these constants
we have assumed that the couplings of these interactions are defined
in terms of ``charges'' (with dimensions of $[E]^{1/2}[L]^{1/2}$). }.
%This is not
%always the adopted convention. For instance the strong force scale is 
%usually represented by the mass of the proton $m_p$, or more concretely 
%by a charge (of dimensions $E^{1/2}L^{1/2}$) $Gm_p^2$. However clearly
%the mass of the proton must be made up in part by the ``electromagnetic''
%energy generated by the color charge, $g_s$. In fact our procedure
%for expressing the strength of the strong interactions is a better
%measurement of the strong interaction coupling 
%because $m_p$ is also made up of standard electromagnetic
%energy.}. 
Suppose further that 
the ratios between the various
constants, $r_{ij}=\alpha_i/\alpha_j$, were observed to be constant.
Choosing what dimensional constants were indeed constants would still
be a matter of taste. 
One could still define a theory in which the various charges
$g_i$ change in time, with fixed ratios, and $\hbar c$ remains constant. 
However it would perhaps start to make more sense, merely
for reasons of simplicity, to postulate instead a changing $\hbar c$.

Therefore, even though a variable $c$ cannot be made a dimensionless
statement, evidence in favour of theoretical models with varying $c$ could be
accrued if the other $\alpha_i$ changed, with fixed ratios.

\section{Cosmological horizons}\label{coshor}
\begin{figure}
\centerline{\psfig{file=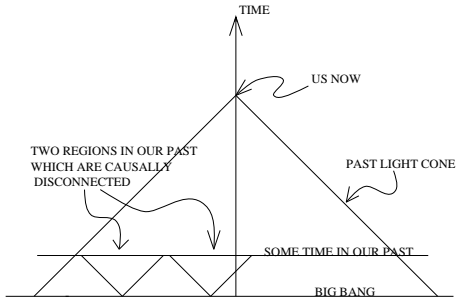,width=6 cm,angle=-90}}
\caption{Conformal diagram (light at $45^\circ$) showing the
horizon structure in the SBB model. Our past light cone contains 
regions outside each others' horizon.}
\label{fig1}
\end{figure}

Perhaps the most puzzling feature of the SBB is the presence 
of cosmological horizons. At any given time any observer
can only see a finite region of the Universe, with comoving radius
$r_h=c\eta$, where $\eta$ denotes conformal time, and 
$c$ the speed of light. Since the horizon
size increases with time we can now observe many regions in our past 
light cone which are causally disconnected, that is, outside each others'
horizon (see Fig.~\ref{fig1}). 
The fact that these regions have the same properties (eg.
Cosmic Microwave background temperatures equal 
to a few parts in $10^5$) is puzzling
as they have not been in physical contact. This is a mystery one may
simply relegate to the setting up of initial conditions in our Universe.

\begin{figure}
\centerline{\psfig{file=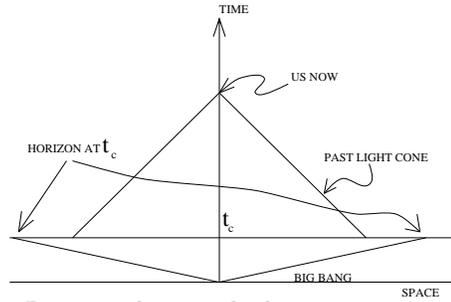,width=6 cm,angle=-90}}
\caption{Diagram showing the horizon structure in a SBB model
in which at time $t_c$ the speed of light changed from $c^-$
to $c^+\ll c^-$. Light travels at $45^\circ$ after $t_c$
but it travels at a much smaller angle with the space axis before
$t_c$. Hence it is possible for the horizon at $t_c$ to be much
larger than the portion of the Universe at $t_c$ intersecting our 
past light cone. All regions in our past have then always been 
in causal contact.}
\label{fig2}
\end{figure}

One may however try to explain these very peculiar initial conditions.
The horizon problem is solved by inflationary scenarios by postulating
a period of accelerated or superluminal 
expansion, that is, if $a$ is the expansion
factor of the Universe, a period with $\ddot a>0$. 
The Friedman equations
require that the strong energy condition $\rho + 3p/c^2 \ge 0$ must then
be violated, where $\rho c^2$ and $p$ are the energy density and pressure
of the cosmic matter. This violation is achieved by the inflaton field. 
If $\ddot a>0$ for a sufficiently long period one can show
that cosmological horizons are a post-inflation 
illusion, and that the whole
observed Universe has in fact been in causal contact
since an early time. 

A more minimalistic
way of solving this problem is to postulate that light
travelled faster in the Early Universe. Suppose there was a ``phase
transition'' at time $t_c$ when the speed of light changed from $c^-$ to
$c^+$. Our past light cone intersects $t=t_c$ at a sphere 
with comoving radius
$r=c^+ (\eta_0-\eta_c)$, where $\eta_0$ and $\eta_c$ are the conformal
times now and at $t_c$. This is as much of the Universe after the
phase transition 
as we can see today~\cite{note2}. On the other hand the horizon size at $t_c$
has comoving radius $r_h =c^-\eta_c$. If $c^-/c^+\gg\eta_0/\eta_c$,
then $r\ll r_h$, meaning that the whole observable Universe today has
in fact always been in causal contact (see Fig.~\ref{fig2}). 
Some simple  manipulations show
that this requires
\begin{equation}\label{cond1}
\log_{10}{c^-\over c^+}\gg 32 -{1\over 2}\log_{10}z_{eq}+{1\over 2}
\log_{10}{T^+_c\over T^+_P}
\end{equation}
where $z_{eq}$ is the redshift at matter radiation equality, and $T^+_c$
and $T^+_P$ are the Universe and the Planck temperatures after the phase
transition. If $T^+_c\approx  T^+_P$ this implies light travelling more
than 30 orders of magnitude faster before the phase transition. 
It is tempting, for symmetry reasons, simply to postulate that 
$c^-=\infty$ but this is not strictly necessary.

\section{A prescription for modifying physical laws while the
speed of light is varying}\label{post}

Hidden in the above argument is the assumption that the 
geometry of the Universe is not affected by a changing $c$.
We have allowed a changing $c$ to do the job normally
done by ``superluminal expansion''. To enhance this effect
we have forced the geometry to still be the SBB geometry. 
We now elaborate on this assumption. 
We will propose a prescription for how, in general, to modify 
gravitational laws while $c$ is changing. This 
prescription is merely the one we found the most fertile.
In Appendix I we describe in detail a theory which 
realizes this prescription. 

The basic assumption is that a variable $c$ does not induce 
corrections to curvature in the cosmological frame, and that
Einstein's equations, relating curvature to stress energy,
are still valid. The rationale behind this postulate is that
$c$ changes in the local Lorentzian frames associated 
with cosmological expansion. The effect
is a special relativistic effect, not a gravitational effect.
Therefore curvature should not feel a changing $c$.

The previous statement is not covariant. However introducing
a function $c(t)$ is not even Lorentz invariant. So it is not
surprising that a favoured gauge, or coordinate choice, must be
made, where the function $c(t)$ is specified, and in which the
above postulate holds true. The cosmological frame
(with the cosmological time $t$) provides such a preferred frame.

In a cosmological setting the postulate proposed implies
that Friedman equations remain valid even when $\dot c\neq 0$:
\begin{eqnarray}
{\left({\dot a\over a}\right)}^2&=&{8\pi G\over 3}\rho -{Kc^2\over a^2}
\label{fried1}\\
{\ddot a\over a}&=&-{4\pi G\over 3}{\left(\rho+3{p\over c^2}\right)}
\label{fried2}
\end{eqnarray}
where, we recall, $\rho c^2$ and $p$ are the energy and 
pressure densities, 
$K=0,\pm 1$ and $G$ the curvature and the gravitational
constants, and the dot denotes a derivative with respect to proper time.
If the Universe is radiation dominated, $p=\rho c^2/3$, and we
have as usual $a\propto t^{1/2}$. 
We have assumed that a frame exists where $c=c(t)$, and identified
this frame with the cosmological frame. 

The assumption that Einstein's equations remain unaffected by 
decelerating light carries with it an important consequence.
Bianchi identities apply to curvature, as a geometrical identity.
These then imply stress energy conservation as an integrability
condition for Einstein's equations. 
If $\dot c\neq 0$,  however,
this integrability condition is not stress energy
conservation. Source terms, proportional to $\dot c/c$,
come about in the conservation equations.

Seen in another way, the conservation equations imply an
equation of motion for free falling point particles. 
This is normally the geodesic equation,
but now source terms will appear in the geodesic equation.
Clearly a violation of the weak equivalence principle is implied
while $c$ is changing \cite{will}. This, of course, does not
conflict with experiment, as we take $\dot c\neq 0$ only in the Early
Universe, possibly for only a very short time (such as a 
phase transition).

Although this is a general remark we shall be concerned mostly
with violations of energy conservation in a cosmological
setting. Friedman equations can be combined into a 
``conservation equation'' with source terms in 
$\dot c/c$ and $\dot G/G$:
\begin{equation}\label{cons1}
\dot\rho+3{\dot a\over a}{\left(\rho+{p\over c^2}\right)}=
-\rho{\dot G\over G}+{3Kc^2\over 4\pi G a^2}{\dot c\over c}
\end{equation}
%or equivalently
%\begin{equation}\label{cons2}
%\dot\rho+3{\dot a\over a}{\left(\rho+{p\over c^2}\right)}=
%\rho{\left(2{\dot c\over c}-{\dot G\over G}\right)}-{3\over 4\pi G}
%{\left({\dot a\over a}\right)}^2{\dot c\over c}
%\end{equation}
In a flat Universe ($K=0$) a changing $c$ does not violate
mass conservation. Energy, on the other hand, 
is proportional to $c^2$. If,
however, $K\neq 0$ not even mass is conserved.

In Eqn.~\ref{cons1} we have included the effects 
of $\dot G$ under the same postulate merely for completness. 
In such a formulation VSL does not reduce to Brans Dicke theory
when $\dot c=0$, and $\dot G\neq 0$. 
This is because 
we postulate that Friedmann equations remain unchanged, 
which implies that the conservation equations acquire terms in 
$\dot c$ and $\dot G$.
In Brans Dicke theory one postulates exactly the opposite: 
the  conservation equations must still be valid, so that the 
weak equivalence principle is satisfied.
While we could have taken this stance 
for $c$ as well we feel that violation of energy conservation is the hallmark 
of changing $c$. Variable $c$ must break Poincare invariance, 
for which energy is the Noether current. Barrow \cite{bdvsl}
has proposed a formulation of VSL which has the correct Brans
Dicke limit.

\section{The flatness puzzle}
We now turn to the flatness puzzle.
The flatness puzzle can be illustrated as follows.
Let $\rho_c$ be the critical density of the Universe:
\begin{equation}
\rho_c={3\over8\pi G}{\left(\dot a\over a\right)}^2
\end{equation}
that is, the mass density corresponding to $K=0$
for a given value of $\dot a/a$. Let us define
$\epsilon=\Omega-1$ with $\Omega=\rho/\rho_c$. Then  
\begin{equation}
\dot\epsilon=(1+\epsilon){\left({\dot\rho\over\rho}-
{\dot\rho_c\over\rho_c}
\right)}
\end{equation}
If $p=w\rho c^2$ (with $\dot w=0$), using 
Eqns.(\ref{fried1}), (\ref{fried2}), and 
(\ref{cons1}) we have:
\begin{eqnarray}
{\dot\rho\over \rho}&=&-3{\dot a\over a}(1+w)-{\dot G\over G}+
2{\dot c\over c}{\epsilon\over 1+\epsilon}\\
{\dot\rho_c\over \rho_c}&=&-{\dot a\over a}(2+(1+\epsilon)(1+3w))
-{\dot G\over G}
\end{eqnarray}
and so
\begin{equation}\label{epsiloneq}
\dot\epsilon=(1+\epsilon)\epsilon {\dot a\over a} 
{\left(1+3w\right)}+2{\dot c\over c}\epsilon
\end{equation}
In the SBB $\epsilon$ grows like $a^2$ in the radiation era, 
and like $a$ in the matter era, leading to a total growth by 
32 orders of magnitude since the Planck epoch. The observational 
fact that $\epsilon$ can at most be of order 1
nowadays requires that either $\epsilon=0$
strictly, or an amazing fine tuning must have existed in the initial
conditions ($\epsilon<10^{-32}$ at $t=t_P$). This is the flatness puzzle.

The $\epsilon=0$ solution is in fact unstable for any matter 
field satisfying the strong energy condition $1+3w>0$. Inflation
solves the flatness problem with an inflaton field which satisfies
$1+3w<0$. For such a field $\epsilon$ is driven towards
zero instead of away from it. Thus inflation can solve the
flatness puzzle.

As Eqn.~\ref{epsiloneq} shows a decreasing speed of light 
($\dot c/c<0$) would also drive $\epsilon$ to 0. If the speed 
of light changes in a sharp phase transition, with $|\dot c/c|\gg
\dot a/a$, we can neglect the expansion terms in 
Eqn.~\ref{epsiloneq}. Then $\dot\epsilon/\epsilon=2\dot c/c$ so
that $\epsilon\propto c^2$. A short calculation shows that the 
condition (\ref{cond1}) also ensures 
that $\epsilon\ll 1$ nowadays, if $\epsilon\approx 1$ before the
transition. 

The instability of the $K\neq 0$ Universes while $\dot c/c<0$ can be
expected simply from inspection of the non conservation equation
Eq.~(\ref{cons1}). Indeed if $\rho$ is above its critical value,
then $K=1$, and Eq.~(\ref{cons1}) tells us that mass is taken out 
of the Universe. If $\rho<\rho_c$, then $K=-1$, and then mass is produced.
Either way the mass density is pushed towards its critical value
$\rho_c$. In contrast with the Big Bang model, during a period 
with $\dot c/c<0$ only the $K=0$ Universe is stable.

Note that with the set of assumptions we have used a changing 
$G$ cannot solve the flatness problem 
(cf.\cite{robert,jana,turner}).

We have assumed in the previous discussion that we are close,
but not fine-tuned, to flatness before the transition. 
It is curious to note that this need not be the case. 
Suppose instead that the Universe acquires ``natural initial
conditions'' (eg. $\epsilon\approx 1$) well
before the phase transition occurs. If such Universes 
are closed they recollapse before the transition. If they are
open, then they approach $\epsilon=-1$.  This is the Milne Universe,
which is our case (constant $G$) may be seen as Minkowski space-time. 
Such a curvature dominated Universe is essentially empty, and a coordinate 
transformation can transform it into Minkowski space-time. Inflation
cannot save these empty Universes, as can be seen from Eqn.~\ref{epsiloneq}.
Indeed even if $1+3w<0$ the first term will be
negligible if $\epsilon\approx-1$. This is not true for VSL: the
second term will still push an $\epsilon=-1$ Universe towards
$\epsilon=0$.

Heuristically this results from the fact that the violations of 
energy conservation responsible for pushing the Universe towards 
flatness do not depend on there being any matter in the Universe.
This can be seen from inspection of Eqn.~(\ref{cons1}).

In this type of scenario it does not matter how far before
the transition the ``initial conditions'' are imposed. We 
end up with a chaotic scenario in which Darwinian selection gets rid
of all the closed Universes. The open Universes become empty and cold.
In the winter of these Universes a phase transition
in $c$ occurs, producing matter, and leaving the Universe
very fine tuned, indeed as an Einstein deSitter Universe (EDSU).

\section{The cosmological constant problem}
There are two types of cosmological constant problems, and
we wish to start our discussion by differentiating them.
Let us write the action as:
\begin{equation}
S=\int dx^4 \sqrt{-g}{\left( {c^4 (R+2\Lambda_1)\over 16\pi G}
 +{\cal L}_M + {\cal L}_{\Lambda_2}\right)}
\end{equation}
where ${\cal L}_M$ is the matter fields Lagragian. 
The term in $\Lambda_1$ is a geometrical cosmological constant,
as first introduced by Einstein. The term in $\Lambda_2$ represents
the vacuum energy density of the quantum fields \cite{steve}.
Both tend to dominate the energy density of the Universe,
leading to the so-called cosmological constant problem.
However they represent two rather different problems.
We shall attempt to solve the problem associated with 
the first, not the second, term.
Ususally one hopes that the second term will be cancelled by an
additional couter-term in the Lagrangian. In the rest of
this paper it is the geometrical cosmological constant
that is under scrutiny.

If the cosmological constant $\Lambda\neq 0$ then the
argument in the previous section
still applies, with $\rho=\rho_m+\rho_\Lambda$,
where $\rho_m$ is the mass density in normal matter, and
\begin{equation}\label{enerlamb}
\rho_\Lambda={\Lambda c^2\over 8\pi G}
\end{equation} 
is the mass density in the cosmological constant.
One still predicts $\Omega_m+\Omega_\Lambda=1$, with
$\Omega_m=\rho_m/\rho_c$ and $\Omega_\Lambda=\rho_\Lambda/\rho_c$.
However now we also have
\begin{equation}\label{dotLm}
\dot\rho_m+3{\dot a\over a}{\left(\rho_m+{p_m\over c^2}
\right)}=-\dot\rho_\Lambda-\rho{\dot G\over G}+
{3K c^2\over 4\pi G a^2}{\dot c\over c}
\end{equation} 
If $\Lambda$ is indeed a constant then from Eq.~(\ref{enerlamb})
\begin{equation}\label{dotL}
{\dot \rho_\Lambda\over \rho_\Lambda}=2{\dot c\over c} -{\dot G
\over G}
\end{equation} 
If we define $\epsilon_\Lambda=\rho_\Lambda/\rho_m$
we then find, after some straightforward algebra, that
\begin{equation}\label{epslab}
\dot \epsilon_\Lambda =\epsilon_\Lambda{\left(
3{\dot a\over a}(1+w)+2{\dot c\over c}{1+\epsilon_\Lambda
\over 1+\epsilon}\right)}
\end{equation} 
Thus, in the SBB model, 
$\epsilon_\Lambda$  increases like $a^4$ in the radiation era, 
like $a^3$ in the matter era,
leading to a total growth by 64 orders of magnitude since the Planck
epoch. 
Again it is puzzling that $\epsilon_\Lambda$ is observationally
known to be at most of order 1
nowadays. We have to face another fine tuning problem in the SBB
model: the cosmological constant problem.

If $\dot c=0$ the solution $\epsilon_\Lambda=0$
is in fact unstable for any $w>-1$. Hence violating the  strong
energy condition $1+3w>0$ would not solve this problem.
Even in the limiting case $w=-1$ the solution 
$\epsilon_\Lambda=0$ is not an attractor: $\epsilon_\Lambda$
would merely remain constant during inflation, then starting to
grow like $a^4$ after inflation.
Therefore inflation cannot ``explain'' the small value
of $\epsilon_{\Lambda}$, as it can with $\epsilon$,
unless one violates the dominant energy condition
$w\ge -1$. 

However, as Eqn.~(\ref{epslab}) shows, 
a period with $\dot c/ c \ll 0$ would drive $ \epsilon_\Lambda$
to zero.  If the speed of light changes suddenly ($|\dot c/c|
\gg \dot a/a$) then we can neglect terms in $\dot a/a$, and so
\begin{equation}
{\dot \epsilon_\Lambda\over \epsilon_\Lambda(1+\epsilon_\Lambda)}
=2{\dot c\over c}{1\over 1+\epsilon}
\end{equation} 
which when combined with $\dot\epsilon/\epsilon=2\dot c/c$
leads to
\begin{equation}
{\epsilon_\Lambda\over 1+\epsilon_\Lambda}
\propto {\epsilon \over 1+\epsilon}
\end{equation} 
The exact constraint on the required 
change in $c$ depends on the initial conditions
in $\epsilon$ and $\epsilon_\Lambda$. In any case once both
$\epsilon\approx 1$ and $\epsilon_\Lambda\approx 1$ we have 
$\epsilon_\Lambda\propto c^2$. Then we can solve the 
cosmological constant problem in a sudden phase transition 
in which
\begin{equation}\label{cond2}
\log_{10}{c^-\over c^+}\gg 64 -{1\over 2}\log_{10}z_{eq}+2\log_{10}
{T^+_c\over T^+_P}
\end{equation}
This condition is considerably more restrictive than (\ref{cond1}),
and means a change in $c$ by more than 60 orders of magnitude,
if $T^+_c\approx  T^+_P$.
Note that once again a period with $\dot G/G$ would not solve
the cosmological constant problem. 

Equations (\ref{epsiloneq}) and (\ref{epslab}) are the equations
one should integrate to find conditions for
 solving the flatness and
cosmological constant problems for arbitrary initial conditions and
with arbitrary curves $c(t)$.  They generalize the conditions 
(\ref{cond1}) and ({\ref{cond2}) which are valid only 
for a starting point with $\epsilon\approx 1$ and
$\epsilon_ \Lambda \approx 1$ and for a step function $c(t)$.

As in the case of the flatness problem we do not need to impose
``natural initial conditions'' ($\epsilon_\Lambda\approx 1$)
just before the transition. These could have existed any time
before the transition, and the argument would still go through,
albeit with a rather different overall picture for the history of the
Universe.

If $\epsilon_\Lambda\approx 1$ well before the transition, then
the Universe soon becomes dominated by the cosmological constant.
We have inflation! The curvature and matter will be inflated away.
We end up in a de-Sitter Universe. When the transition is about to occur
it finds a flat Universe ($\epsilon=0$), with no matter ($\rho_m=0$),
and with a cosmological constant. If we rewrite Eqn.(\ref{epslab})
in terms of $\epsilon_m=\rho_m/\rho_\Lambda$, for $\epsilon=0$
and $|\dot c/c|\gg \dot a /a$, we have 
$\dot \epsilon_m=-2(\dot c/ c)(1+\epsilon_m)$. Integrating
leads to $1+\epsilon_m\propto c^{-2}$.
We conclude that we do not need the presence of any matter in the Universe
for a VSL transition to convert a cosmological
constant dominated Universe into a EDSU Universe full of ordinary
matter. This can be seen
from Eqns.~(\ref{dotLm})-(\ref{dotL}). A sharp decline in $c$ will
always discharge any vacuum energy density into ordinary matter.

We stress the curious point that 
in this type of scenario the flatness problem is not
solved by VSL, but rather by the period of inflation
preceding VSL.

% [It is perhaps distracting, but the fate of anti-Desitter is now
% the same as closed Universes before: they get selected out]
%{\tt Bit about anti-desitter removed}

%Even in the sudden phase transition case there are
%two complications to the simple picture in \cite{us}. First
%the dynamical relevance of the cosmological constant decreases slower
%than naively expected while the curvature $K$ is still relevant.
%On the other hand the energy density in the cosmological constant
%is dumped into the radiation, accelerating the process in
%which $\Lambda$ becomes dynamically irrelevant. The exact
%constraint on the required 
%change in $c$ depends on the initial conditions
%in $\epsilon$ and $\epsilon_\Lambda$. In any case once both
%$\epsilon\ll1$ and $\epsilon_\Lambda\ll 1$ we have 

\section{The homogeneity of the Universe}\label{homo}
Solving the horizon problem by no means guarantees solving 
the homogeneity problem, that is, the uncanny homogeneity of 
the currently observed Universe
across many regions which have apparently been causally disconnected.
Although solving the horizon problem is a necessary condition for solving
the homogeneity problem, in a generic inflationary model solving the 
first causes serious 
problems in solving the latter.  Early causal contact between
the entire observed Universe allows
equilibration processes to homogenize the whole observed
Universe.  It is crucial to the inflation picture that before
inflation the observable universe in well inside the Jeans length,
and thus equilibrates toward a homogeneous state.
However no such process is perfect, and small density
fluctuations tend to be left outside the Hubble radius, 
once the Universe resumes its
standard Big Bang course. These fluctuations then grow like $a^2$
during the radiation era, like $a$ during the matter era, usually entailing
a very inhomogeneous Universe nowadays. This is a common flaw in
early inflationary models \cite{gupi} which requires additional
fine-tuning to resolve.

In order to approach this problem we study in Appendix II the 
effects of a changing $c$ on the theory of scalar cosmological 
perturbations \cite{KS}. The basic result is that the comoving
density contrast $\Delta$ and gauge-invariant velocity $v$
are subject to the equations:
\begin{eqnarray}
\Delta'-{\left(3w{a'\over a}+{c'\over c}\right)}\Delta&=&
-(1+w)kv-2{a'\over a}w\Pi_T\label{delcdotm}\\
v'+{\left({a'\over a}-2{c'\over c}\right)}v&=&{\left(
{c_s^2 k\over 1+w} -{3\over 2k}
{a'\over a}{\left({a'\over a}+{c'\over c}\right)}
\right)}\Delta \nonumber \\
+{kc^2w\over  1+w}\Gamma-
&kc&{\left({2/3\over 1+w}+{3\over k^2c^2}
{\left(a'\over a\right)}^2\right)}w\Pi_T\label{vcdotm}
\end{eqnarray}
where $k$ is the wave vector of the fluctuations,
and $\Gamma$ is the entropy production rate, $\Pi_T$
the anisotropic stress, and $c_s$ the speed of sound,
according to definitions spelled out in Appendix II.

In the case of a sudden phase transition Eqn.~({\ref{delcdotm})
shows us that $\Delta\propto c$, regardless of the chosen 
equations of state for $\Gamma$ and $\Pi_T$. Hence 
\begin{equation}
{\Delta^+\over\Delta^-}={c^+\over c^-}
\end{equation}
meaning a suppression of any fluctuations before the phase
transition by more than a factor of $10^{-60}$ if condition
(\ref{cond2}) is satisfied.
The suppression of fluctuations induced by a sudden phase transition
in $c$ can be intuitively understood in the same fashion as the solution to
the flatness problem. Mass conservation violation
ensures that only a Universe at critical mass density is stable,
if $\dot c/c\ll 0$. But this process occurs locally, so
after the phase transition the Universe should be left
at critical density {\it locally}. Hence the suppression of 
density fluctuations.

We next need to know what are the initial conditions for $\Delta$ and
$v$. Suppose that at some very early time $t_i$
one has $\dot c/c= 0$ and the whole observable Universe
nowadays is inside the Jeans length: $\eta_0\ll c_i\eta_i/{\sqrt 3}$.
The latter condition is enforced as a byproduct of solving the horizon
problem. The whole observable Universe nowadays is then
initially in a thermal state. What is more each portion of the Universe
can be described by the canonical ensemble and so the Universe
is homogeneous apart from thermal fluctuations \cite{Peebles}.
These are characterized by the mass fluctuation 
\begin{equation}
\sigma^{2}_M={{\langle\delta M ^2
\rangle}\over{\langle M\rangle}^2}={4k_b T_i\over M c_i^2}
\end{equation}
Converted into a power spectrum for $\Delta$ this is a 
white noise spectrum with amplitude
\begin{equation}\label{pdelta}
P_\Delta(k)={\langle |\Delta(k)^2|\rangle}\propto 
{4k_bT_i\over \rho_ic_i^2}
\end{equation}

What happens to a thermal distribution, its temperature, and its
fluctuations, while $c$ is changing?
%In order to answer this question we must distinguish between two
%aspects of temperature: color and effective temperature.
%Color temperature refers to the distribution function of particle energies,
%which is the Planck distribution $P(E)=1/(e^{E/k_bT_c}-1)$, 
%where $T_c$ is the color temperature. 
In thermal equilibrium the distribution function of particle energies
is the Planck distribution $P(E)=1/(e^{E/k_bT}-1)$, where $T$
is the temperature. 
When one integrates over the whole phase space, one obtains
the bulk energy density $\rho c^2\propto (k_b T)^4/(\hbar c)^3$.
%where $T_e$ is the effective temperature. For a black body spectrum
%$T_e=T_c$, but in general $T_e\leq T_c$ (called a gray body).  {\\tt
%needs more work}
Let us now consider the time when the Universe has already 
flattened out sufficiently for mass to be approximately 
conserved. To define the situation more completely, we 
make two additional microphysical assumptions.
Firstly, let mass be conserved also for individual quantum particles,
so that their energies scale like $E\propto c^2$. 
Secondly,  we assume particles' wavelengths do not change with $c$. 
If homogeneity is preserved, indeed the wavelength is an 
adiabatic invariant, fixed by a set of quantum numbers, 
eg: $\lambda =L/n$ for a particle in a box of size $L$.
 
Under the first of these assumptions a Planckian distribution with
temperature $T$ remains Planckian, but $T\propto c^2$.
Under the second assumption, we have $\lambda=2\pi \hbar c/E$, 
and so $\hbar/ c$ should remain constant. Therefore the phase space 
structure is changed so that, without particle production, one still
has $\rho c^2\propto (k_b T)^4/(\hbar c)^3$, with $T\propto c^2$.
%the
%effective temperature is still equal to the color temperature,
%and changes like $c^2$. Indeed from $\rho c^2\propto (k_b T_e)^4/
%(\hbar c)^3$ we have $T_e\propto c^2$.
A black body therefore remains a black body,
with a temperature $T\propto c^2$. If we combine this effect 
with expansion, with the aid of Eqn.~(\ref{cons1}) we have
\begin{equation}\label{temp}
\dot T + T{\left({\dot a\over a}-2{\dot c\over c}\right)}=0
\end{equation}
We can then integrate this equation through the epoch when
$c$ is changing to find the temperature $T_i$ of the initial
state. This fully fixes the initial conditions for scalar
fluctuations, by means of (\ref{pdelta}). 

In the case of a sudden phase transition we have $T^+=
T^- c^{2+}/c^{2-}$, and so 
\begin{equation}
\sigma_M^{2-}={4k_b T^-\over M c^{2-}}={4k_b T^+\over M c^{2+}}
\end{equation}
or 
\begin{equation}
\Delta^-(k)^{2}\approx {4k_bT^+\over \rho^+ c^{2+}}
\end{equation}
but since $\Delta\propto c$ we have 
\begin{equation}
\Delta^+(k)\approx {\sqrt {4k_bT^+\over \rho^+ c^{2+}}}{c^+\over c^-}
\end{equation}
Even if $T^+=T_P^+=10^{19}Gev$ these fluctuations would still be 
negligible nowadays. Therefore although the Universe ends up in a 
thermal state after the phase transition, its thermal fluctuations,
associated with the canonical ensemble, are strongly suppressed.

For a more general $c(t)$ function the procedure is as follows.
Integrate Eqn.~(\ref{temp}) backwards up to a time $t_i$ when
$\dot c=0$, to find $T(t_i)$. Give $\Delta(t_i)$ a thermal spectrum of 
fluctuations, according to (\ref{pdelta}), with $T(t_i)$. 
With this initial condition integrate Eqns.(\ref{delcdotm})  
and  (\ref{vcdotm}) (or even better the second order equation 
(\ref{deltacddot}) given in Appendix II), to find
$\Delta$ nowadays. 

It is conceivable that a careful design of $c(t)$
would leave fluctuations, once $\dot c=0$ again,
with the right amplitude and spectrum to explain structure formation.
In particular $c(t)$ may be designed so as to convert a white noise
spectrum into a scale-invariant spectrum. However we feel that 
until a mechanism for inducing $c(t)$ is found such efforts
are bound to look ludicrously contrived.

We feel that the power of VSL scenarios is precisely in leaving the
Universe very homogenous, after $c$ has stopped changing. This would 
then set the stage for causal mechanisms of structure formation
to do their job \cite{vs,aa}.

\section{The isotropy of the Universe}
There is a sense in which there is an isotropy problem
in the SBB model, similar to the homogeneity problem.
We follow closely the remark made in \cite{KS}, pp.26. 

In Appendix~III we write down the vector Einstein's equations 
in the vector gauge,
and from them we derive the vorticity ``conservation'' equation when
$\dot c/c\neq 0$. If $v$ is the vorticity (defined in Appendix) and
$\Pi^T$ the vector stress, we have:
\begin{equation}
v'+(1-3w){a'\over a}v-2{c'\over c}v=-{kc\over 2}{w\over 1+w}\Pi^T
\end{equation}
In the absence of driving stress, $v$
remains constant during the radiation dominated epoch, and 
decays like
$1/a$ in the matter epoch. In \cite{KS} it is further argued 
that the relevant dimensionless quantity is
\begin{equation}
\omega={(k/ a)v \over(a'/ ca)}\propto {1\over a^{(1-9w)/ 2}}
\end{equation}
Hence for $w>1/9$
vorticity grows, leading to a further fine tuning problem.

This is most notably a problem if we accept the Planck
equipartition proposal, introduced in \cite{barrow}.
At Planck epoch there would then be a significant
vorticity. Depending on how one looks at it, this vorticity
would then get frozen in or grow, leading to a very anisotropic
Universe nowadays.

Whether or not this is a problem is clearly debatable.
In any case either inflation or VSL models could 
solve this prospective problem. For $w<-1/3$ we have that 
$v$ decays faster than $1/a^2$. Whatever dimensionless
quantity one chooses to look at, vorticity is therefore
safely inflated away. If $\dot c/c\neq 0$ we have that 
$v\propto c^2$. Again any primordial vorticity is safely 
suppressed after a phase transition in $c$ satisfying 
any of the conditions (\ref{cond1}) or (\ref{cond2}).

\section{The entropy problem and setting the initial conditions}
Let us first consider the SBB model.
Let $S_h$ be the entropy inside the horizon, and $\sigma_h=S_h/k_B$
be its dimensionless counterpart. $\sigma_h$ is of order
$10^{96}$ nowadays. If we assume that the only scales in the cosmological
model are the ones provided by the fundamental constants, then at $t_P$ the 
temperature is $T_P$. At Planck time, $\sigma_h$ (being dimensionless)
is naturally of order 1. In the SBB model the horizon distance is $d_h=2t$
in the radiation dominated epoch,
and ignoring mass thresholds $t\propto 1/T^2$. If evolution is adiabatic
one then has (in a flat Universe)
\begin{equation}\label{hbbs}
\sigma_h(t)\approx\sigma_h(t_P){\left(T_P\over T\right)}^3.
\end{equation}
Since $\sigma_h(t_P)\sim 1$, one has $\sigma_h(t_0)\sim 10^{96}$.
Thus the large entropy inside the horizon nowadays is a reflection of 
the lack of scales beyond the ones provided by
the fundamental constants, the fact that the horizon size
is much larger nowadays than at Planck time, and the flatness of the Universe. 
One may rephrase the horizon and flatness problems in terms of entropy 
\cite{infl1}. However if one is willing to accept the horizon 
structure and flatness of the Universe simply as features of the initial 
conditions (rather than problems), there is no additional entropy problem.

There {\em is} a problem that arises if one tries to solve the horizon problem,
keeping the adiabatic assumption, by means of superluminal expansion.
This blows what at Planck time
is a region much smaller than the Planck size into a comoving
region containing the whole observable Universe nowadays.
This solves the horizon problem. However
if evolution is adiabatic such a process implies that
$\sigma_h(t_0)\ll 1$. Stated in another way, 
since the number of particles inside the horizon $n_h$ is of the same
order as $\sigma_h$, this implies an empty Universe nowadays.

More mathematically, if $d_h$ is the horizon proper distance,
one has
\begin{equation}\label{entropy}
{\dot\sigma_h\over \sigma_h}={3\over d_h}
\end{equation}
where we have used $d_h=a\int^tdt'/a$. With any standard matter
($p>-\rho c^2/3$) the horizon grows like $t$. Accordingly $\sigma_h$
grows like a power of $t$. 
On the other hand
 the horizon grows faster than $t$ if $p<-\rho c^2/3$:
it grows exponentially if $p=-\rho c^2$, and like $t^n$ (with $n>1$)
for $-\rho c^2<p<-\rho c^2/3$. This provides the inflationary solution to
the horizon problem. However in the latter case  Eqn. (\ref{entropy})
implies that $\sigma_h$ decreases exponentially, leading to $\sigma_h(t_0)
\ll 1$. 
The way inflation bypasses this problem is by dropping the adiabatic
assumption. Indeed during inflation the Universe supercools, and a period
of reheating follows the end of inflation\footnote{
This issue has been carefully analyzed in the context of
inflationary models and models with time varying $G$ in \cite{turner}
}.

In a VSL scenarios the detailed solution to the entropy problem 
depends on when and what type of ``natural conditions''
are given to the pre transition Universe. 
We first derive equations for the entropy under varying $c$. 
From $s=(4/3)\rho c^2/T$, $\rho\propto T^4/(\hbar c)^3$, and
from Eqns~\ref{cons1} and \ref{dotLm}
we obtain that the entropy of radiation satisfies
\begin{equation}\label{dots}
{\dot s\over s}={3\over 4}{\dot\rho\over\rho}
= -3{\dot a\over a} +{3\over 2}{\dot c\over c}{\epsilon
(1+\epsilon_\Lambda)
\over 1+\epsilon}-{3\over 2}{\dot c\over c}\epsilon_\Lambda
\end{equation}
%and its temperature
%\begin{equation}
%{\dot T\over T}= -{\dot a\over a} + 2{\dot c\over c}+
%{1\over 2}{\dot c\over c}{\epsilon
%(1+\epsilon_\Lambda)
%\over 1+\epsilon}-{1\over 2}{\dot c\over c}\epsilon_\Lambda
%\end{equation}
If the Universe is EDSU, there are no violations of mass conservation,
and entropy is conserved. However if the Universe is open or has a positive
cosmological constant, then we have seen that there is creation of mass.
Accordingly there must be creation of particles, and entropy is produced.
If the Universe is closed, particles are taken away, and the entropy decreases.

The most suspicious case is therefore if the Universe was Einstein
de-Sitter before the phase transition. Let us assume therefore that
at $t=t_P^-$ (the Planck time with the constants before the transition)
the entropy inside the horizon (which has proper size $c^-t_P^-$)
was of order 1. Then the entropy inside the Hubble volume at $t=t_P^+$,
before and after the transition, is
\begin{equation}
\sigma_h(t_P^+)=\sigma_h(t_P^-)
{\left( c^+t_P^+\over c^- t_P^-\right)}^3{\left(
a(t_P^-)\over a(t_P^+)\right)}^3\approx 1
\end{equation}
where we have used $t_P^+/t_P^-=(c^-/c^+)^2$. 
One takes a fraction $(c^+/c^-)^3$
of the horizon volume before the transition to make the Hubble 
volume after the transition. However the entropy inside the horizon
has increased since $t_P^+$ by the same factor. Therefore entropy
conservation in this case does not conflict with $\sigma_h(t_P^+)
\approx 1$ after the transition.  One way of understanding this is that by 
imposing flatness from the outset (before the transition) one has already 
``solved'' the entropy problem. Notice that the above argument
works for any value of $t_c/t_P^+$.

Now consider the case where ``natural'' initial conditions were 
also imposed at $t_P^-$, with $\Lambda=0$.  One should
have $\epsilon(t_P^-)$ of order 1. We have already discussed how the flatness 
problem is solved in this case, when large empty curvature dominated 
universes are filled with a (nearly perfectly) critical energy density 
during the transition. 
Open Universes become very empty, but they are still pushed
to EDSU at the transition. One may integrate (\ref{dots}) to find
that $s^+/s^-=(1+\epsilon^-)^{-3/4}$. One may also use Eqn.~\ref{epsiloneq}
to find that $\epsilon$ has evolved since $t_P^-$ to 
$\epsilon^-(t_P^+)+1\approx (a(t_P^-)/a(t_P^+))^2\approx (t_P^-
/t_P^+)^2$, where we have used $a\propto t$ for the Milne Universe.
Hence we have that during the transition entropy is produced
like $s^+/s^-=(t_P^+/t_P^-)^{3/2}=(c^-/c^+)^3$. Given that $a\propto t$
for such Universes, the entropy before the transition in the
proper volume of size $c^+t_P^+$ is
\begin{equation}
S^-(c^+t_P^+)=
{\left( c^+t_P^+\over c^- t_P^-\right)}^3{\left(
a(t_P^-)\over a(t_P^+)\right)}^3\approx {\left( c^+\over c^-\right)}^3
\end{equation}
that is there is practically no entropy in relevant volume before
the transition. However we have that after the transition
\begin{equation}
\sigma_h(t_P^+)=S^+(c^+t_P^+)=
{S^-(c^+t_P^+)\left( c^-\over c^+\right)}^3\approx 1
\end{equation}
In such scenarios the Universe is rather cold and empty before
the transition. However the transition itself reheats the Universe. 
Notice that, like in the first case discussed, the above argument
works for any value of $t_c/t_P^+$.

If at $t=t_P^-$ one also has $\epsilon_\Lambda\approx 1$ then we have a
scenario in which the cosmological constant dominates, solves
the flatness problem, and is discharged into normal matter. However
if $\rho_\Lambda\approx\rho_P^-$ at $t\approx t_P^-$, 
then whatever the transition
time, after the transition the Universe will have a density in normal
matter equal to $\rho_m=\rho_P^-$. Hence the Hubble time after the 
transition will be $t_P^-$, whatever the actual age of the Universe.
One may integrate (\ref{dots}) to find that in this case (setting
$\epsilon=0$) the entropy production during the transition is 
$s^+/s^-=(1+\epsilon_\Lambda^-)^{3/4}$. In the period between
$t=t_P^-$ and the transition, $\epsilon_\Lambda$
increases like $a^4$, and the entropy density is  diluted
like $1/a^3$. Hence after the transition the entropy density is 
what it was at $t=t_P^-$, that is $s^+\approx 1/L_P^{-3}$. 
If we now follow the Universe
until its Hubble time is $t_P^+$ (when its density is $\rho_P^+$)
we must wait until the expansion factor has increased by a factor
of $(\rho_P^+/\rho_P^-)^{1/4}$. Given that $s\propto 1/a^3$ 
the entropy density is diluted by a factor of $(\rho_P^+/\rho_P^-)^{3/4}$.
Therefore the entropy density when the Hubble time is $t=t_P^+$
is $s\approx 1/L_P^{+3}$. Again the dimensionless entropy inside the
Hubble volume, when this has size $L_P^+$, is of order 1.

%The upshot is that whatever effective 
%energy density is in the cosmological constant before the trasition gets 
%translated into an energy density in matter after the transition boosted by a 
%factor XXX.  If we set up a ``natural'' cosmological constant at  
%$t_P^-$, we get a $\rho_+$ many times $\rho_P^+$, an apparantly awakward 
%result. 

Finally it is worth noting that treating the pre-transition universe 
simply as a Roberston-Walker model is no doubt overly simplistic, and we use 
it simply as a device to introducing our ideas.  We expect that further 
development of these ideas could result in a radically different view of the 
pre-transition phase (much as has happened with the inflationary scenario).  
One interesting observation is that one could avoid having multiple Planck
times by considering that $G\propto c^4$. Such assumption would
not conflict with the dynamics of flatness and $\Lambda$, as shown before,
but now $t_P^-=t_P^+$.

\section{Conclusions}
We have shown how a time varying speed of light could
provide a resolution to the well known cosmological puzzles.   These
``VSL'' models could provide an alternative to the standard
Inflationary picture, and furthermore 
resolve the classical cosmological constant puzzle.  
At a technical level, the proposed VSL picture is not nearly as
well developed as the inflationary one, and one purpose of this
article is to stimulate further work on the unresolved technical
issues.  
We are not trying to take an ``anti-inflation'' stand, but we do
strongly feel that broadening the range of possible models of the
very early Universe would be very healthy for the field of cosmology,
and would ultimately allow us to state in more concrete terms the
extent to which one model is preferred. 

On a more fundamental level we hope to expand the phenomenological
approach presented in this paper into a theory where the concept
of (Poincare) symmetry breaking provides the physical basis for
VSL. Symmetry breaking is also the central ingredient in 
causal theories of structure formation. We therefore hope to
arrive at a scenario where symmetry breaking provides a complete
and consistent complement to the SBB model which can resolve the
standard puzzles as well as explain the origin of cosmic structure.

{\bf Note added in proof:}
  
After this paper, as well as its sequel \cite{sq1}, were completed
J. Moffat brought to our attention
two papers in which he proposes a similar idea \cite{mof}. While
Moffat's work does not go as far as ours in addressing the 
flatness, cosmological constant, and entropy problems, he does 
go considerably further than we have in terms of specific
model building.  Moffat's model does not satisfy our
prescription for solving the cosmological problems, but it may do so
in a modified form.  We are currently investigating this possibility.
We regret that because we were unaware of this work we did not cite it
in the first publicly distributed version of this paper.

\section*{Acknowledgements} We would like to thank John Barrow,
Ruth Durrer, Robert
Brandenberger, Gary Gibbons, and Erick Weinberg for discussion. 
We acknowledge support from
PPARC (A.A.) and the Royal Society (J.M.).

\section*{Appendix I: A specific realization of VSL}
In this Appendix we set up a specific VSL theory. 
We first discuss the simple case of the electrodynamics of the 
point particle in Minkowski space time. We start from Bekenstein's
theory of variable $\alpha$, and show how a VSL alternative 
could be set up. We highlight the subtleties encountered in the VSL
formulation. We then
perform the same exercise with the Einstein-Hilbert action.
We briefly consider the
dynamics of the field $\psi=c^4$. Finally we cast the 
key elements of our construction into a body of axioms.

\subsection{Electrodynamics in flat space time}
A changing $\alpha$ theory was proposed by Bekenstein
\cite{bek2} based on the
postulate of Lorentz invariance. The electrodynamics of a point
particle was first analyzed. If Lorentz invariance
is to be preserved then the particle mass $m$ and its charge 
$e$ must be variable. In order to preserve ``minimal coupling'' (reduction
to standard electromagnetism when $\alpha={\rm const}$) one chooses
the world line action 
\begin{equation}\label{emp}
L=-mc{\sqrt{-u^\mu u_\mu}} +{e\over c}u^\mu A_\mu
\end{equation}
with $u^\mu=\dot x^\mu$, $g_{\mu\nu}=\eta_{\mu\nu}$, 
$e=e(x^\mu)$, and $m=m(x^\mu)$. Minimal coupling means simply
to take the standard action and  replace $e$ and $m$ by variables
without breaking Lorentz invariance. $e$ and $m$ must then be scalar
functions.
This action leads to equation:
\begin{equation}
(m\dot x_\mu\dot)=-m_{,\mu}c^2 +{e\over c} u^\nu F_{\mu\nu}
\end{equation}
with the electromagnetic field tensor defined as 
\begin{equation}
F_{\mu\nu}={1\over e}(\partial_\mu(eA_\nu )-\partial_\nu(eA_\mu ))
\end{equation}
The electromagnetic action can therefore be defined as:
\begin{equation}
S_{EM}={-1\over 16\pi}\int d^4 x F_{\mu\nu}F^{\mu\nu}
\end{equation}
Also, the particle action ($\ref{emp}$) may be written 
as a Lagrangian density:
\begin{equation}
S_M=\int d^4x {\delta^{(3)}({\bf x}-{\bf x}(\tau))\over \gamma}
(-mc^2+(e/c)u^\mu A_\mu)
\end{equation}
in which $\gamma$ is the Lorentz factor. 
Maxwell's equations are then:
\begin{equation}
e\partial_\mu(F^{\mu\nu}/e)=4\pi j^\mu
\end{equation}
with the current
\begin{equation}
j^\mu={\delta^{(3)}({\bf x}-{\bf x}(\tau))\over \gamma}{eu^\mu\over c}
\end{equation}
This current\footnote{
There is an alternative view in which rather than a changing $e$
one considers that the vacuum is a dielectric medium with variable
$\epsilon$. One may then identify a conserved charge, but this
is not the charge which couples to the gauge field.}
is the current which couples to the gauge field,
and in the rest frame it equals $e$. Therefore it cannot be conserved,
and indeed we have that 
\begin{equation}
\partial_\mu j^\mu={j^\mu\over e^2}\partial_\mu e 
\end{equation}

Let us now postulate instead that a changing $\alpha$ is to be interpreted
as $c\propto\hbar\propto\alpha^{-1/2}$, and that $e$ and $m$ are to be seen
as constants. Minimal coupling, in the above sense, would then prompt
us to consider the action ($\ref{emp}$), but with $c=c(x^\mu)$ everywhere,
and $e$ and $m$ constants. This action leads
to equations:
\begin{equation}\label{eq1}
m\ddot x_\mu={1\over 2}(mc^2)_{,\mu}+ {e\over c} u^\nu F_{\mu\nu}
\end{equation}
with the electromagnetic tensor defined as 
\begin{equation}\label{eq2}
F_{\mu\nu}=c(\partial_\mu(A_\nu/c )-\partial_\nu(A_\mu/c ))
\end{equation}

However the above construction is not complete. 
In spite of the appearance of Eqns.~($\ref{emp}$), ($\ref{eq1}$),
and ($\ref{eq2}$), Lorentz invariance is broken. This boils down
to the fact that, say $\partial_\mu$ is no long a 4-vector. Even
if $c$ were to be regarded as a scalar, $\partial_\mu$ would contain
$c$ in its zero component, but not in its spatial components. 
The usual contractions leading to $S$  could still be taken
but $S$ would no longer be a scalar. This manifests itself in
the equations ($\ref{eq1}$) in the fact that in $\ddot x^\mu$ 
there are terms in $\partial c$ which
break Lorentz invariance.

Since the action is not Lorentz invariant, a minimal coupling prescription
cannot possibly be true in every coordinate system. 
Minimal coupling is now the statement that there
is a preferred reference frame in which the action is to be obtained from
the standard action simply by replacing $c$ with a field. Let us call this
frame the ``light frame''.
In regions in which $c$ changes very little changes in the action upon
Lorentz transformations are negligible. Hence all boosts performed upon
the light frame become nearly equivalent and Lorentz invariance is recovered.

The Maxwell equations in a VSL theory become
\begin{equation}\label{max2}
{1\over c}\partial_\mu(cF^{\mu\nu})=4\pi j^\mu
\end{equation}
in the light frame.
Given that Lorentz invariance is broken, one can no longer expect the
general expression for a conserved current to take the form
$\partial_\mu j^\mu=0$.
Indeed one could try and compute $\partial_\nu$ of equations (\ref{max2}),
but now $\partial_\mu$ and $\partial_\nu$ do not commute. Also their
commutator is not Lorentz invariant: for instance $[\partial_0,
\partial_i]=(-\partial_i c/c^2)\partial_0$. 
Still, $\partial_\mu j^\mu=0$ holds in the time frame.  It is just
that this expression transforms into something more complicated in
other frames.  The more complicated expression would still place
constraints on the theory, which could still be called ``conservation
of charge''.

\subsection{Minimal coupling to gravity}
Let us now examine gravity in such a theory\footnote{
Gravitation is normally
regarded as the gauge theory of the Poincare group\cite{tomk}. 
Here we simply abandon
this point of view. In some future work we will try to define
a gauge principle for broken symmetries, thereby recovering the
standard view}. As in the previous
case we will impose a minimal coupling principle. Working in
analogy with Brans-Dicke theory, let us define a field $\psi=c^4$,
and introduce the following action
\begin{equation}\label{s}
S=\int dx^4{\left( \sqrt{-g}{\left( {\psi (R+2\Lambda)\over 16\pi G}
 +{\cal L}_M\right)} +{\cal L}_\psi \right)}
\end{equation}
The dynamical variables are a metric $g_{\mu\nu}$, any matter field
variables contained in ${\cal L}_M$, and $\psi$ itself.
The Riemann tensor (and the Ricci scalar) is to be computed from
$g_{\mu\nu}$ at constant $\psi$ in the usual way. 

As in the previous section covariance is broken, in spite of all
appearances. $\psi$ does not appear in coordinate transformations 
of the metric, and so the connection $\Gamma^\alpha_{\mu\nu}$ does not
contain terms in $\nabla \psi$ in any frame. However the connection will
contain different terms in $\psi$ in different frames. Hence the statement
that the Riemann tensor is to be computed from the metric at constant 
$\psi$ can only be true in one preferred frame. Minimal coupling
requires the definition of a light frame. The action (\ref{s}) is only 
Lorentz invariant in appearance.

Varying the action with respect to the metric leads to:
\begin{eqnarray}\label{var}
{\delta S\over \delta g^{\mu\nu}}&=&{\sqrt{-g}\psi\over 8\pi G}
[G_{\mu\nu}-g_{\mu\nu}\Lambda]\\
{\delta S_M\over \delta g^{\mu\nu}}&=&-{\sqrt{-g}\psi\over 8\pi G}
T_{\mu\nu}
\end{eqnarray}
leading to a set of Einstein's equations without any extra terms
\begin{equation}
G_{\mu\nu}-g_{\mu\nu}\Lambda = {8\pi G\over \psi} T_{\mu\nu}
\end{equation}
valid in the light frame. This is the way we chose to phrase our
postulates in SectionIII of our paper. In other words all we need
is minimal coupling at the level of Einstein's equations.

The fact that a favoured set of coordinates is picked by our action
principle is not surprising as Lorentz invariance is broken. On
the other hand notice that the dielectric vacuum of Bekenstein theory
is an ether theory. His theory also breaks Lorentz invariance, not at the level
of the laws of physics, but in the form of the contents of space-time.

In changing $\alpha$ theories a favoured frame is always picked up. 
In a cosmological setting it makes sense to identify this frame with
the cosmological frame. Free falling observers comoving with the 
cosmological flow define a proper time and a set of spatial coordinates,
to be identified with the light frame. In this frame the Einstein-Hilbert
action is minimally coupled to a changing $\psi$, and the same happens to
Friedmann equations. The rest of our paper follows.

\subsection{The dynamics of $\psi$}
The definition of ${\cal L}_\psi$ controls the dynamics of $\psi$.
This is the most speculative aspect of our theory, but it also opens
the doors to empirical model building. In our paper we preferred
a scenario in which $c$ changes in an abrupt phase transition, but
one could also imagine $c\propto a^n$.
The latter scenario would result from a Brans Dicke type of Lagrangian
\begin{equation}
{\cal L}_\psi={-\omega\over 16\pi G\psi}\dot\psi^2 
\end{equation}
(where $\omega$ is a dimensionless coupling)
and is being investigated. Addition of a temperature dependent
potential $V(\psi)$ would induce a phase transition, as in the scenario
developped in our paper.

However here we only make the following remarks, which are independent
of any concrete choice of ${\cal L}_\psi$. If $K=\Lambda=0$ one has
\begin{equation}
{\delta {\cal L}_\psi\over \delta \psi}={\sqrt{-g} T\over 4\psi}
\end{equation}
and so in the radiation dominated epoch ($T=0$), 
once $K=\Lambda=0$, one should
not expect driving terms for the $\psi$ equation. Hence once the cosmological
problems are solved, in the radiation epoch, $c$ and $\hbar$
should be constants.
Incidently, once 
the matter dominated epoch ($T\neq0$) is reached, $\psi$ should
perhaps start changing again, with interesting observations 
consequences \cite{webb}.
We are studying the phase space portraits of these cosmologies, when
say $\Lambda\neq0$, and with various ${\cal L}_\psi$.

During phase transitions the perfect fluid approximation must break down.
One should then use, say, scalar field theory (let's call it $\phi$). 
Now notice that terms in $\dot\phi$ will act as a source to $\psi$
(as they contain the speed of light). Hence whenever there is a
phase transition and the VEV of a field changes a large amount,
one may expect a large change in the speed of light, with most choices
of ${\cal L}_\psi$. A changing $\psi$ associated with SSB could then solve
the quantum version of the cosmological constant problem, but this might
require a rather contorted choice of ${\cal L}_\psi$.

\subsection{Axiomatic formulation of VSL theories}

{\bf Postulate 1.} {\it A changing $\alpha$ is to be interpreted as
a changing $c$ and $\hbar$ in the ratios $c\propto\hbar\propto
\alpha^{-1/2}$. The coupling $e$ is constant.}

This postulate merely sets up the theoretical interpretation of 
the possible experimental fact that $\alpha$ changes, in terms of variable
dimensional quantities. This is a matter of convention and not
experiment, as much as
a constant $\hbar c$ is a matter of convention. With the above choice
a system of units for mass, length, time, and temperature is unambiguously
defined. 

{\bf Postulate 2.} {\it There is a preferred frame for the laws of physics.
This preferred frame is normally suggested by the symmetries of the 
problem, or by a criterium such as $c=c(t)$.}

If $c$ is variable, Lorentz invariance must be broken. Even if one
writes Lorentz invariant looking expressions these do not transform
covariantly. [In general this boils down to the explicit presence of $c$ 
in the operator $\partial_\mu$. Once one admits that Lorentz invariance 
must be explicitly broken then a preferred frame must exist to formulate
the laws of physics. These laws are not invariant under frame 
transformation, and one may expect that a preferred frame exists
where these laws simplify.

{\bf Postulate 3.} {\it In the preferred frame one may obtain the laws of
physics simply by replacing $c$ in the standard (Lorentz invariant) 
action, wherever it occurs,
by a field $c=c(x^\mu)$.}

This is the principle of minimal coupling. Because the laws of physics
cannot be Lorentz invariant it 
will not hold in every frame. 
Hence the
application of this postulate depends crucially on the previous postulate
supplying us with a favoured frame. This principle may apply in Minkowski
space time electrodynamics, scalar field theory, etc, in which case 
the frame in which $c=c(t)$ is probably the best choice. The cosmological
frame, endowed with the cosmic proper time is probably the best choice
in a cosmological setting.

{\bf Postulate 4} {\it The dynamics of $c$ must be determined by 
an action principle deriving from adding an extra term to the Lagrangian
which is a function of $c$ only.}

This is work in progress. We do not wish to specify this postulate further
because for all we know this extra term can be anything. We merely
specify that no fields (including the metric) must be present in this
extra term because we wish minimal coupling to propagate into
the Einstein's equations.

\section*{Appendix II: Scalar perturbation equations for VSL models}
In this Appendix we derive the scalar cosmological perturbation equations 
in VSL scenarios. 
We assume $K=\Lambda=0$, and use a gauge where the perturbed metric
is written as
\begin{equation}
ds^2=a^2[-(1+2AY)d\eta^2 -2BYk_idx^id\eta +\delta_{ij}dx^idx^j]
\end{equation}
for a Fourier component with wave vector $k^i$. Here $Y$ is
a scalar harmonic. 
We shall use conformal time $\eta$ to study fluctuations,
and denote $'=d/d\eta$. 
The stress energy tensor is also writen as
\begin{eqnarray}
\delta T^0_0&=&-\rho Y\delta\nonumber\\
\delta T^i_0&=&-{\left(\rho+{p\over c^2}\right)}{v\over c}k^iY
\nonumber\\
\delta T^i_j&=&p\Pi_LY\delta^i_j+(k^ik_j-1/3\delta^i_j k^2)Yp\Pi_T
\end{eqnarray}
The Einstein's constraint equations then read \cite{KS}
\begin{eqnarray}
{3\over c^2}{\left(a'\over a\right)}^2A
-{1\over c}{a'\over a}kB&=&-{4\pi Ga^2\over c^2}\rho\delta\\
{k\over c}{a'\over a}A -{\left( {\left(a'\over a\right)}'-
{\left(a'\over a\right)}^2\right)}{B\over c^2}&=&
{4\pi Ga^2\over c^2}{\left(\rho+{p\over c^2}\right)}{v\over c}
\end{eqnarray}
and the dynamical equations are
\begin{eqnarray}
A+{1\over kc}{\left(B'+2{a'\over a}B\right)}&=&
-{8\pi Ga^2\over c^2}{p\over c^2}{\Pi_T\over k^2}\\
{a'\over a}{A'\over c^2}+{\left( 2{a''\over a}-
{\left(a'\over a\right)}^2\right)}{A\over c^2}&=&
{4\pi Ga^2\over c^2}{p\over c^2}{\left(\Pi_L-2\Pi_T\right)}
\end{eqnarray}
We assume that these equations do not receive corrections
in $\dot c/c$. This statement is gauge-dependent, much like
its counterpart for the unperturbed Eintein's equations.
We can only hope that the physical result does not change
qualitatively from gauge to gauge. Complying
with tradition we now define the comoving density contrast
\cite{KS}
\begin{equation}
\Delta=\delta+3(1+w){a'\over ca}{1\over k}{\left(
{v\over c}-B\right)}
\end{equation}
We also introduce the entropy production rate
\begin{equation}
\Gamma=\Pi_L-{c_s^2\over wc^2}\delta
\end{equation}
where the speed of sound $c_s$ is given by
\begin{equation}\label{cs}
c^2_s={p'\over \rho '}=wc^2{\left(1-{2\over 3}{1\over 1+w}
{c'\over c}{a\over a'}\right)}
\end{equation}
Note that the thermodynamical speed of sound is given by
$c^2_s=(\partial p/\partial \rho)|_S$. Since in SBB models
evolution is isentropic  $c^2_s=(\partial p/\partial \rho)|_S
=\dot p/\dot\rho=p'/\rho '$. When $\dot c\neq 0$ evolution need not be
isentropic. However we keep the definition $c_s^2=p'/\rho '$
since this is the definition used in perturbative calculations.
One must however remember that the speed of sound given in
(\ref{cs}) is not the usual thermodynamical quantity. 
With this definition one has for adiabatic perturbations
$\delta p/\delta\rho=p'/\rho'$, that is the ratio between
pressure and density fluctuations mimics the ratio of their 
background rates of change.

Combining all four Einstein's equations we can then obtain
the (non-)conservation equations \cite{KS}
\begin{eqnarray}
\Delta'-{\left(3w{a'\over a}+{c'\over c}\right)}\Delta&=&
-(1+w)kv-2{a'\over a}w\Pi_T\label{delcdot2}\\
v'+{\left({a'\over a}-2{c'\over c}\right)}v&=&{\left(
{c_s^2 k\over 1+w} -{3\over 2k}
{a'\over a}{\left({a'\over a}+{c'\over c}\right)}
\right)}\Delta \nonumber \\
+{kc^2w\over  1+w}\Gamma-
&kc&{\left({2/3\over 1+w}+{3\over k^2c^2}
{\left(a'\over a\right)}^2\right)}w\Pi_T\label{vcdot2}
\end{eqnarray}
These can then be combined into a second order equation for
$\Delta$. If $\Gamma=\Pi_T=0$ this equation takes the form
\begin{equation}\label{deltacddot}
\Delta''+f\Delta'+(g+h+c_s^2k^2)\Delta=0
\end{equation}
with 
\begin{eqnarray}
f&=&(1-3w)-3{c'\over c}\\
g&=&{\left(a'\over a\right)}^2{\left(\frac{9}{2}
w^2-3w-\frac{3}{2}\right)}\\
h&=&2{\left(c'\over c\right)}^2+{\left(\frac{9w}{2}-\frac{5}{2}
\right)}{c'\over c}{a'\over a}-{\left(c'\over c\right)}'
\end{eqnarray}

\section*{Appendix III: Vector perturbation equations for VSL models}
In a similar fashion we can study vector modes in a gauge
where the metric may be written as
\begin{equation}
ds^2=a^2[-d\eta^2 +2B Y_i dx^id\eta +\delta_{ij}dx^idx^j]
\end{equation}
where $Y_i$ is a vector harmonic. The stress energy tensor 
is written as
\begin{eqnarray}
\delta T^0_i&=&{\left(\rho+{p\over c^2}\right)}{\left({v\over
c}-B\right)}Y_i\\
\delta T_{ij}&=&p\Pi^TY_{(i,j)}
\end{eqnarray}
Einstein's equations then read \cite{KS}
\begin{eqnarray}
k^2B&=&{16\pi G\over c^2}a^2{\left(\rho+{p\over c^2}\right)}
{v\over c}\\
{k\over c}{\left( B'+2{a'\over a}B\right)}&=&-{8\pi G\over c^2}
{p\over c^2}a^2\Pi^T
\end{eqnarray}
We assume that these do not receive $\dot c/ c$ corrections.
The conservation equation is then:
\begin{equation}
v'+(1-3w){a'\over a}v-2{c'\over c}v=-{kc\over 2}{w\over 1+w}\Pi^T
\end{equation}


\begin{thebibliography}{99}
\bibitem{note2} Of course if $c^-=\infty$ we could in fact see the whole
Universe before $t_c$. This also happens for a deSitter initial phase,
where the event horizon diverges at the BigBang.
\end{thebibliography}
\end{document}